Predicting Long-term Outcomes of Educational Interventions Using the Evolutionary Causal

Matrices and Markov Chain Based on Educational Neuroscience


Hyemin Han[a]*

University of Alabama

Kangwook Lee[b]          Firat Soylu[a]

KAIST          University of Alabama

Author Note

[a]Educational Psychology Program, University of Alabama, Box 870231, Tuscaloosa, AL 35487, USA.

[b]School of Electrical Engineering, KAIST, 291 Daehak-ro, Yeseong-gu, Daejeon 34141, Republic of Korea.

* Correspondence concerning this article should be addressed to Hyemin Han, Educational Psychology Program, University of Alabama, Tuscaloosa, Box 870231, AL 35487. Email: hyemin.han@ua.edu. Phone: 1-250-348-0746. Fax: 1-205-348-0683.








**Abstract**

We developed a prediction model based on the evolutionary causal matrices (ECM) and the Markov Chain to predict long-term influences of educational interventions on adolescents' development. Particularly, we created a computational model predicting longitudinal influences of different types of stories of moral exemplars on adolescents' voluntary service participation. We tested whether the developed prediction model can properly predict a long-term longitudinal trend of change in voluntary service participation rate by comparing prediction results and surveyed data. Furthermore, we examined which type of intervention would most effectively promote service engagement and what is the minimum required frequency of intervention to produce a large effect. We discussed the implications of the developed prediction model in educational interventions based on educational neuroscience.

## Introduction

Yoda: It is the future you see.

Luke: The future? Will they die?

Yoda: Difficult to see. Always in motion is the future.

*- Star Wars: Episode V - The Empire Strikes Back*

Psychological intervention experiments in educational settings have been conducted to examine how to enhance positive youth development, such as academic adjustment and well-being, among adolescents [1]. Recently, psychologists have developed various intervention methods and tested their longitudinal influences on diverse domains, including but not limited to, adolescents' academic motivation, belongingness to school contexts and social competences to deal with bullying issues in school settings [2–5]. These intervention studies potentially contribute to the improvement of school environment and finally adolescents' development based on empirical evidence [6,7]. However, because they have tested effects of interventions in experimental settings, which are decontextualized, more restricted, controlled and involve a smaller sample compared to real school settings, it would be difficult to directly apply developed interventions to classroom contexts [8]. Thus, large-scale, long-term longitudinal studies examining diverse intervention methods adopted in school curricular and activities should be conducted to overcome this shortcoming. For instance, researchers should investigate which type of intervention can effectively promote developmental change and how often it should be conducted in classrooms in order to produce a significant and large effect. By answering these questions, educators and educational policy makers can better understand how to properly apply psychological interventions to enhance the quality of education in real school settings at the



macroscopic level, such as the district level. However, it would be difficult to test the long-term effects of different types and frequencies of interventions with real adolescent populations due to limited time and resources [9].

The meta-analysis is perhaps a feasible and reliable way to examine which intervention methods can properly work in general by systematically reviewing and analyzing various methods developed in multiple studies [10]. Several scholars have conducted meta-analyses to systematically review the effectiveness of diverse educational interventions in diverse contexts [6,8,11,12]. They have identified which types of intervention programs can produce a significant effect on positive youth development [13]. However, they were not able to provide complete answers to questions that educators and policy makers may raise. For instance, although those previous meta-analyses of interventions examined the effect size of each type of interventions, they could not provide any information needed to determine the minimum frequency of interventions required to produce a significant and large effect. Moreover, the majority of previous meta-analysis studies were mainly interested in demonstrating whether or not a certain type of intervention can produce a significant effect overall [6,12], instead of more directly examining which type of intervention can be more effective than others. Thus, meta-analysis itself would not be sufficient to provide practical implications to educators and policy makers.

We intend to employ the framework of evolutionary and computational theory and develop a computational model to examine which type of intervention is effective and how often it should be conducted to produce a significant and large effect in the long term. We predict future long-term outcomes based on relatively small-scale, short-term data gathered from lab and classroom experiments. Evolutionary theory provides the present study with a theoretical scaffold to approach the current problem, that is, the long-term prediction of intervention outcomes, in a



practical manner. Based on ideas in evolutionary theory, particularly the evolutionary causal matrices (ECM), we establish an evolutionary model modeling how a system consisting of adolescents will evolve over time while being influenced by interventions [14–16]. In order to implement the ECM-based longitudinal prediction model, we employed Markov Chain analysis [17]. Finally, we developed a simulation program based on the computational model to test whether the model properly predicts longitudinal outcomes.

In short, the present study aims to predict long-term outcomes of educational interventions using relatively small-scale, short-term data by applying the ideas of the ECM and Markov Chain. The developed computational model will be able to provide useful insights about how to apply intervention models to diverse educational settings to educators and policy makers. Furthermore, based on the developed computational model and prediction findings, the present study discusses their implications for future educational intervention studies in educational psychology. Particularly, we focus on how this computational approach can contribute to the improvement of interventions based on an interdisciplinary theoretical framework incorporating perspectives from neuroscience, cognitive science, and education.

**Evolutionary Causal Matrices**

The ECM that was inspired by the theory of biological evolution provide useful tools to predict the ratio of a certain type of individuals among the whole population and where the equilibrium point will be under certain selection pressures in the long term, particularly for the studies of cultural evolution [15]. ECM consist of multiple matrices describing the dynamics in certain systems. Each matrix in ECM describes the probability of the longitudinal transition between certain states [16]. For instance, we may consider a simple illustrative example of the longitudinal transition between conformers and non-conformers in cultural systems. Conformers



can be defined as a group of individuals who conform to certain social norms; while, non-conformers do not observe the norms. There are two different types of systems, with different selection pressures. The first cultural system ($C_1$) is well organized and has plentiful resources available to individuals. In this system, conformers are more likely to have better fitness compared to non-conformers, and non-conformers are likely to follow the social norms over time. The second cultural system ($C_2$) does not have enough resources to support individuals and is not well organized. Non-conformers are more likely to be successful in this system. ECM describing the transition between two states from $t$ to $t+1$ are presented in Table 1. We can calculate the ratio of each state in the $C_1$ at $t+1$ from data at $t$ as follows:

*<Place Table 1 Here>*

80% of initial conformers will still be conformers while 20% of them will become non-conformers at $t+1$.

60% of initial non-conformers will still be non-conformers while 40% of them will become conformers at $t+1$.

The ratio can also be calculated in case of the $C_2$ similarly. In general, we can calculate the ratio of state A in the whole population at t+1 using this formula [14,15]:

$$F_A(t+1) = \frac{\sum_{i \in P} F_i(t) ECM_{iA}}{\sum_{i \in P} \left[ F_i(t) \sum_{j \in p} ECM_{ij} \right]} \qquad (1)$$

We can show that those two systems eventually reach a certain equilibrium in the long term using the ECM and the formula. For instance, let's say there are 50% of conformers and 50% of non-conformers in both the $C_1$ and $C_2$ at $t = 0$. After conducting iterative calculations, both systems reach an equilibrium. On the one hand, in case of the $C_1$, the ratio of conformers to non-



conformers converges to 75:25 at $t = 10$. On the other hand, that ratio converges to 87.5:12.5 at $t = 10$ in the case of the $C_2$.

This methodology can also be applied to predicting outcomes of interventions in groups, which can be regarded as systems. Interventions change the dynamics in a certain system and finally each state. Previous intervention studies have demonstrated that interventions altered group norms, influenced the dynamics within as well as between individuals in the group, and finally changed the individuals' behavior [18–20]. Therefore, we can create ECM based on findings from intervention experiments informing longitudinal changes between various behavioral states. One matrix is created per intervention type. In each matrix, a number in each cell is calculated by the transition rate from a certain state, which can be represented by a different type of behavior, at $t$ to another or same state at $t$+1. Using the created ECM, future intervention outcomes can be predicted by iteratively calculating the ratio of each state among the whole population at a certain time point.

**Markov Chain**

Markov chain is a mathematical tool, which can be used to model and analyze stochastic systems [17]. It has been widely used in a wide range of fields in science and engineering. One of the most famous, recent successes of Markov chain is Google's PageRank [21]: modeling behavior of web surfers as a Markov chain, Google's PageRank efficiently ranks an enormous number of web pages on the Internet based on search metrics.

We provide a formal definition of Markov chains as follows. A Markov chain is defined with a set of states, transition matrix, and initial state distribution; in this work, we consider only discrete time Markov chains. At $t = 0$, the state of a Markov chain is in its initial, denoted by $S_0$, and this initial state is randomly chosen according to the initial state distribution. The transition





matrix dictates how a Markov chain (randomly) evolves. When the number of states is $p$, the size of the transition matrix is $p$ by $p$, and each row of the transition matrix defines how the Markov chain evolves from each state. More precisely, if the state of the chain at time $t$, denoted by $S_t$, is $x$, $1 \leq x \leq p$, the state of the chain at time $t + 1$, denoted by $S_{t+1}$, is a random variable, and the distribution of this random variable is specified by the $x$-th row of the transition matrix.

The Markov chain defined above is a *time-independent* Markov chain since its random behavior is independent of time, and it depends only on its state. A time-dependent Markov chain is defined with a series of transition matrices, each of which defines the behavior of the chain at each time. That is, at time $t$, the Markov chain behaves according to the $t$-th transition matrix.

The effects of ECM and interventions can be effectively modeled, analyzed, and simulated with an appropriately defined time-dependent Markov chain. That is, a student's engagement is modeled as a binary state (0 if not engaged, 1 if engaged), and the initial state distribution is specified based on empirical data, collected from field experiments. Similarly, for each intervention type (including 'no intervention'), we construct a transition matrix with the empirical data. Finally, we construct an appropriate time series of transition matrices given the type and frequency of interventions. For instance, if the transition matrix corresponding to 'no intervention' is $P_{no}$, the transition matrix corresponding to a certain intervention is $Q$, and the intervention frequency is one per 5 months, the corresponding series of transition matrices is ($P_{no}$, $P_{no}$, $P_{no}$, $P_{no}$, $Q$, $P_{no}$, $P_{no}$, $P_{no}$, $Q$, …).

Once a Markov chain is specified, one can find the probability that a system is in state $x$ at time $t$ for any $x$ and $t$ as follows. Given the initial distribution and the first transition matrix $P$, one can find the probability that the distribution of $S_1$, the state of the system at time 1, as follows:



$$\Pr(S_1 = x) = \sum_{s=1}^{p} P_{s,x} \Pr(S_0 = s). \tag{2}$$

This simple equation, a special case of the Chapman-Kolmogorov forward equation [17], allows us to efficiently analyze the probability distribution of the state at any time $t$. Equivalently, this will allow us to predict what ratio of the students will be in a certain state at time $t$.

## Material and Methods

### Dataset

We used data collected from the two previous intervention experiments [22,23] to create the ECM, which were implemented by the Markov Chain for prediction and simulation. These two intervention experiments compared the effectiveness of attainable and relevant exemplary moral stories, such as moral stories of peer exemplars, and that of extraordinary exemplary stories, such as the biography of Martin Luther King or Eleanor Roosevelt, on the promotion of voluntary service engagement among adolescents.

The first intervention experiment was conducted by recruiting 54 college students (Mean age = 22.17 years, $SD$ = 5.20 years). They were randomly assigned to one of these three conditions: (1) attainable and relevant exemplar intervention, (2) extraordinary exemplar intervention, and (3) control conditions (17, 18 and 19 subjects assigned respectively). The first group was presented with the stories of attainable voluntary service engagement among subjects' peer college students. The extraordinary group was presented with the stories of unattainable extreme service engagement. This experiment presented non-moral stories, such as sports news, to the control group. The pre-test survey was conducted before presenting intervention materials and measured subjects' initial voluntary service engagement. About 1.5 months after the intervention session, the post-test survey was conducted to examine the longitudinal change in service engagement.



The second intervention experiment was conducted at a middle school. 184 8[th] graders (equivalent to 14 years old) were recruited and they were randomly assigned to one of these three groups, similar to the previous experiment: peer exemplar, extraordinary exemplar and control story groups (55, 52 and 77 subjects respectively). The overall experimental procedure was similar to that of the previous experiment. This experiment collected pre-test survey data before the beginning of the intervention session and post-test survey data about 1.5 month after the end of the session, respectively. Consequently, data collected from a total of 238 subjects was included in the dataset for modeling.

**Procedures**

   **Simulation model development.** In order to test whether the established computational model properly predicts long-term intervention outcomes, we implemented the model in a simulation program. First, we created the ECM to describe the number of individuals in each condition at $t+1$ based on data at $t$. The surveyed voluntary service engagement was coded into a binary variable (participated vs. not participated). Then we calculated how many subjects in each participation status at $t$ transitioned to each participation status at $t+1$ for each intervention condition using the dataset. For instance, at the pre-test survey, 40 participants and 32 non-participants were assigned to the attainable and relevant exemplar condition. In this condition, out of 40 initial participants, 36 participants continued to participate while four other participants withdrew from voluntary service at $t+1$. Since the time interval between pre- and post-test surveys in the previous intervention experiments was 1.5 month, we set the time interval in this model ($\Delta t$) as 1.5 months. Out of 32 initial non-participants, 14 started service engagement while 18 continued not to participate at $t+1$. We can summarize these longitudinal transitions as follows:



Running head: PREDICTING EDUCATIONAL INTERVENTION OUTCOMES

(In the attainable and relevant exemplary condition)

Participants ($t$+1) = .90 Participants ($t$) + .44 Non-participants ($t$)

Non-participants ($t$+1) = .10 Participants ($t$) + .56 Non-participants ($t$)

As a result, we can complete the whole ECM using the summarized information.

Second, we implemented the created ECM using the Markov Chain algorithm and composed a MATLAB code for simulation. We created diagrams demonstrating the longitudinal changes between participation statuses in each condition in the form of chain (see Figure 1). We then coded a MATLAB simulation program implementing these Markov Chains. The simulation program was designed to iteratively calculate the number of participants and non-participants at $t$+1 based on data at $t$. $t$+1 data was calculated according to the formula (1). As a result, these steps were implemented in a simulation program as the following pseudo code presents:

*<Place Figure 1 Here>*

```
Sub iterative_simulaion()
 For i = 1 to # of different intervention types (or conditions)
  For j = 1 to # of different frequencies
  participants (1) = initial_participants
  non_participants (1) = initial_non_participants
   For t = 2 to # of iterations
   If t is divisible by j Then # Interventions occur at current t
    participants (t) = participants (t-1) * ECM (i, 1, 1) + non_partcipants (t-1) * ECM (i, 1, 2)
    non_participants (t) = non_participants (t-1) x ECM (i, 2, 1) + participants (t-1) x ECM (i, 2, 2)
   Else # Interventions do not occur at current t
    participants (t) = participants (t-1) * ECM (3, 1, 1)  + non_partcipants (t-1) * ECM (3, 1, 2)
    non_participants (t) = non_participants (t-1) * ECM (3, 2, 1) + participants (t-1) * ECM (3, 2, 2)
   End If
   record_current_status () # Store current statuses in memory
  Next t
```



Next j

Next i

End Sub

The full MATLAB code is available as a supplementary material for reference.

Finally, we added additional routines to visualize simulation results. These routines were composed to plot the differences in the mean participation rates between different conditions by different frequencies, the longitudinal trajectories of participation rates of different conditions by different frequencies, and results of statistical analyses, i.e., ANOVA and comparisons between different conditions and frequencies.

**Predictability test.** Before applying the created simulation model for predicting longitudinal outcomes of interventions, whether the model can properly predict future outcomes should be tested. Thus, we employed large datasets regarding the ratio of voluntary service participants collected by previous surveys with a total of 24,863 subjects [24–26]. Table 3 summarizes the number of voluntary service participants and non-participants in each survey. Because these surveys did not employ any psychological intervention, we compared the simulated ratio of voluntary service participants in the control condition and the ratio in the collected dataset. We set the number of the whole population as 238, which was identical to the number of total subjects in two previous intervention experiments. Among them, 127 were set as participants and 111 were set as non-participants at the initial time point ($t = 0$). After 100 iterations, which is equivalent to $t = 150$ months, we compared the simulated outcome with the real dataset using the chi-squared test method.

*<Place Table 2 here>*



**Simulation result analysis.** Before analyzing results from iterative simulations, for primary analysis, the present study examined whether different types of interventions differentially promoted voluntary service activity in the two previous experiments. We set post-intervention voluntary service engagement as a dependent variable; the engagement was coded into a binary variable (participated vs. not participated). We controlled for the pre-test engagement during analysis.

The longitudinal outcomes of interventions were predicted using the created simulation model. For each intervention condition (attainable and relevant, extraordinary and control), the ratio of voluntary service participants and non-participants among the whole population were calculated until the $100^{th}$ generation, which is equivalent to 150 months after $t = 0$. We set such a relatively long-term time frame for the simulation and analysis, since even a brief psychological intervention might influence students' achievement and social adjustment long-term, at least for a couple of years [5]. Moreover, we tested different longitudinal trajectories of service engagement produced by different frequencies of intervention. We simulated longitudinal trajectories for 50 different frequencies, ranging from once per 1.5 months to once per 75 months. We considered such an extreme case in terms of the frequency, once per 75 months, since previous psychological intervention studies have shown that even a one-time brief intervention was able to produce significant long-term outcomes [5,27]. Similar to the case of the comparison with real data, we entered 238 as the number of the whole population at $t = 0$; likewise, we set 127 as initial voluntary service participants and 111 as initial non-participants.

We conducted linear regression analysis to examine whether the intervention type and frequency significantly influenced the longitudinal trajectories of service engagement rate. We set the intervention type, frequency, and $t$ as independent variables. Furthermore, in order to identify the



best type of intervention for promoting service engagement and the minimum frequency of intervention that is required to produce a significant and large effect, we conducted a series of regression analyses. Moreover, for each frequency, we compared the mean participation ratio between experimental and control conditions. We calculated the mean participation ratio by averaging the participation ratio from $t = 0$ to $t = 100$ for each condition. The partial $\eta^2$ value of each comparison was also calculated to examine the minimum frequency required to produce a practically significant outcome through interventions.

We note that when interventions of the same type are periodically applied to a student group, the number of engaged students will converge to a periodic sequence. For instance, consider the case of exemplar interventions. The number of engaged students surges when an exemplar intervention is applied but immediately starts decaying until it surges again with the next intervention. Hence, the maximum number of engaged students in each cycle converges to a certain number within at most a few cycles. (See Figure 3.) The same argument holds for the number of disengaged students as well. We thus consider the mean participation ratio as a dependent variable for the main analysis since the evolution process is captured in it.

In addition, we examined the predicted maximum and minimum participation rate for each condition and intervention frequency. The intervention type and frequency of interventions were entered to the regression model as independent variables, and either predicted maximum or minimum participation rate was entered to the model as a dependent variable. Similar to the previous regression analysis, we estimated the maximum and minimum participation rate for 50 different frequencies, ranged from once per 1.5 months to once per 75 months.

**Results**



Running head: PREDICTING EDUCATIONAL INTERVENTION OUTCOMES

**Preliminary analysis**

We compared the effectiveness of the attainable and relevant exemplar intervention and the

extraordinary exemplar intervention in the previous experiments. The result of logistic regression

analysis demonstrated that the attainable and the relevant exemplar intervention better promoted

service participation compared to both the extraordinary exemplar intervention, $\beta = 1.44$, $z =$

3.03, $p = .002$, 95% CI [.51 2.38], and the control condition, $\beta = .89$, $z = 2.02$, $p = .04$, 95% CI

[.03 1.76]. However, the longitudinal outcome was not significantly different between the

extraordinary exemplar intervention and control conditions, $\beta = .55$, $z = 1.35$, $p = .18$, 95% CI [-

.25 1.35]. The overall regression model was statistically significant, $\chi^2 (3) = 46.10$, $p < .001$,

Pseudo $R^2 = .19$. Given these findings, an intervention program using attainable and relevant

exemplars better promoted voluntary service engagement compared to extraordinary exemplars.

**Predictability test**

*<Place Table 3 here>*

*<Place Figure 2 here>*

We used the chi-squared test to examine if the created simulation model can properly predict

future outcomes. When we compared the ratio of voluntary service participants between the

simulation result and actual survey result, there was not any statistically significant difference

found between those two, $\chi^2 (1) = 1.53$, $p = .22$, $V = .01$ (see Table 4). Predicted participation

rate in a certain $t$ calculated from the simulation model is presented in Figure 2. Thus, the

simulation model in the present study is deemed to reliably predict the longitudinal change in

voluntary service participants, given that the outcome of simulation did not significantly differ

from the real data.



**Simulation results**

After performing iterative simulations, we tested whether the effects of different intervention types and frequencies were statistically significant in predicting outcomes quantified by the ratio of service participants among the whole population. The results of the regression analysis demonstrated that the effect of intervention type was significant; compared to the control condition, the attainable and relevant exemplar intervention condition significantly better promoted voluntary service engagement, $B = .025$, $t = 27.53$, $p < .001$, 95% CI [.023 .027], while the extraordinary exemplar intervention condition showed a significantly worse result, $B = -.020$, $t = -22.42$, $p < .001$, 95% CI [-.022 -.019]. Moreover, the effect of intervention frequency was significant, $B = -.00013$, $t = -5.18$, $p < .001$, 95% CI [-.00018 -.00008]. The overall model was also statistically significant, $F (4, 14795) = 635.87$, $p < .001$, adjusted $\underline{R^2} = .15$. However, the effect of $t$ was insignificant, $B = -.00$, $t = -.13$, $p = .90$, 95% CI [-.00 .00].

Thus, our simulation model was able to demonstrate significant differences in the predicted long-term outcomes of interventions between different intervention types and frequencies. This finding suggests that attainable and relevant exemplars can better promote service engagement compared to extraordinary exemplars. In addition, the effect of interventions became greater as the interventions were conducted more frequently.

*<Place Figure 3 Here>*

Figure 3 demonstrates differentiated trajectories of longitudinal change in voluntary service engagement between different intervention types and frequencies. Similar to the findings from the two previous intervention experiments, the application of the attainable and relevant exemplar intervention better promoted the mean voluntary service participation rate compared to the control condition; meanwhile, the extraordinary exemplar intervention backfired.



Furthermore, as interventions performed more frequently, the longitudinal trajectory of each experimental condition became more deviated from the trajectory of the control condition over time. For instance, on the one hand, when interventions were performed once per three or six months, the trajectories of experimental conditions did not completely rebound to the baseline in the control condition. On the other hand, when interventions were conducted less frequently (e.g., once per 12 or 24 months), the trajectories of experimental conditions completely converged to the baseline in the control condition during inter-intervention periods.

*<Place Figure 4 Here>*

Moreover, we tested how often the attainable and relevant exemplar interventions should be conducted to produce a significantly large promotion effect compared to the control condition. First, we conducted linear regression analysis to examine the effect of frequency (see Figures 4a and 4b). The effect of interventions became smaller as they were conducted less frequently. Second, we examined the minimum frequency that is required to produce a statistically significant difference in the participation rate between the attainable and relevant exemplar and control conditions. For this comparison, we applied Bonferroni's correction for multiple comparisons. Since three comparisons were involved in this analysis, i.e., control condition vs. attainable and relevant exemplar condition, control vs. unattainable and irrelevant exemplar condition, attainable and relevant exemplar condition vs. unattainable and irrelevant exemplar condition, it would be required to apply such a multiple comparison correction method to use a valid $p$-value threshold. After the application of the Bonferroni's correction, the simulated result reported that the intervention should be conducted at least once per 49.5 months to produce a statistically significant difference between conditions (see Figure 4c). Third, we examined the minimum frequency required to produce a large effect when the mean participation rate was



compared between conditions by calculating the partial $\eta^2$ value. The simulation result demonstrated that interventions should be performed at least once per 10.5 months in order to produce a large effect, which was represented by a partial $\eta^2$ value larger than .14 (see Figure 4d). Given these findings, attainable and relevant exemplars should be presented to individuals at least once per 10.5 months to statistically as well as practically significantly promote voluntary service participation rate in the long term.

*<Place Figure 5 here>*

Finally, we analyzed the maximum and minimum participation rate for each intervention condition and intervention frequency. First, as shown in Figure 5, in the case of the maximum participation rate, the attainable and relevant condition demonstrated the highest maximum participation rate for all possible intervention frequencies among three different conditions. The result of regression analysis also indicated that the attainable and relevant exemplar condition showed a significantly higher predicted maximum participation rate compared to other conditions, $\beta = .99$, $t(149) = 61.69$, $p < .001$, $\eta^2 = .73$. The overall regression model was also significant, $F(3, 149) = 1693.56$, $p < .001$, adjusted $R^2 = .97$. Meanwhile, the predicted minimum participation in the attainable and relevant exemplar condition was significantly greater than that in the unattainable and irrelevant exemplar condition, $\beta = .97$, $t(149) = 38.17$, $p < .001$, $\eta^2 = .70$. However, the difference between the attainable and relevant exemplar condition and the control condition was insignificant, $\beta = .01$, $t(149) = 0.41$, $p = .68$, $\eta^2 = .00$. The overall model was significant, $F(3, 149) = 642.40$, $p < .001$, adjusted $R^2 = .93$.

**Mathematical analysis**

The current prediction model, which has been implemented in the form of a simulation, can also be mathematically analyzed, when the interventions of the same type are regularly applied. In



Figure 3, one may observe that without intervention, the ratio of students engaged in voluntary

service activity converges to 50.87%. This limit can be analytically found by solving an

equilibrium equation. In equilibrium, the ratio of engaged students, denoted by $x$, and the ratio of

unengaged students, or $1 - x$, will remain the same. Thus, we have $0.29(1 - x) = 0.28x$ or $x =$

$\frac{29}{57} \cong 0.5087$. One can also find the equilibrium point of any intervention scheme. If the

attainable and relevant exemplar intervention is applied every month, the equilibrium will be the

solution of $0.44(1 - x) = 0.10x$ or $x = \frac{44}{54} \cong 0.8148$. Similarly, if the extraordinary exemplar

intervention is applied every month, the equilibrium will be the solution of $0.12(1 - x) = 0.36x$

or $x = \frac{12}{48} = 0.2500$. Since intervention is applied intermittently, one would expect that the ratio

of engaged students oscillates between the equilibrium ratio of 'without intervention' and that of

the applied intervention, and this oscillation behavior is clearly observed in Figure 3.

Another observation is that when a periodic intervention is applied, the long-term evolution

converges to a certain periodic pattern. For instance, the ratio of engaged students increases

when an attainable and relevant exemplar intervention is applied, then decreases until the next

intervention is applied, and this periodic pattern repeats. A more delicate analysis can precisely

characterize such periodic patterns. Consider the case where the attainable and relevant exemplar

intervention is applied every $T$ time intervals. That is, the ratio of engaged students evolves

according to the Markov chain corresponding to 'without any intervention' $T - 1$ times in a row,

and then every $T^{\text{th}}$ evolution is according to that of 'attainable and relevant exemplar

intervention'. During the $T - 1$ evolutions, the ratio of engaged students approaches the

equilibrium ratio exponentially fast, and the exponent is the difference between the probability of



'remain as engaged' and that of 'becomes engaged', or $0.72 - 0.29 = 0.43$ [17]. Putting all together, we have the following equations:

$$\left(b - \frac{29}{57}\right) = \left(a - \frac{29}{57}\right)(0.43)^{T-1}, \ a = 0.44 + 0.46b. \tag{3}$$

By solving these equations, one can find the upper peak of the periodic pattern ($a$) and the lower peak of the periodic pattern ($b$). For instance, when the intervention is applied every 3 months, or $T = 2$, we have $b = 0.5973, a = 0.7147$. When the intervention is applied every 6 months, or $T = 4$, we have $b = 0.52241, a = 0.6803$. Note that these analytical results exactly match the simulation results shown in Figure 3.

However, these analytical results could not be applied to cases when the frequency of interventions becomes irregular, or different types of interventions are administrated to the same group. For instance, we may consider the case of Korean moral education curriculum, which requires elementary, middle, and high schoolers to take moral education classes for different numbers of hours per week, e.g., once or twice a week, in order to accommodate each individual school's academic schedule [28]. Moreover, moral educators may attempt to present different types of exemplars, peer as well as historic moral exemplars, simultaneously or in sequence. In these kinds of complicated, but potentially possible cases, we might not have analytical results to predict long-term outcomes. Although such complicated cases were not discussed in the present study, still, computer simulations might have to be conducted to deal with more complicated cases possibly existing in the reality, even though such analytic results are available for some simpler cases.

## Discussion



We developed a computer simulation model predicting long-term outcomes of interventions, particularly those aiming to promote moral development, based on the ideas of the ECM and Markov Chain. The simulation model was able to predict that attainable and relevant exemplars would more effectively promote voluntary service engagement compared to extraordinary exemplar or non-moral stories given the findings from the linear regression analyses of simulation results that we conducted. Moreover, this model showed that the intervention should be conducted at least once per 10.5 months to make a statistically and practically significant difference in the participation rate between the attainable and relevant exemplar condition and control condition.

This study shows how computational models and simulations can be used to explore long-term effectiveness of intervention studies in educational psychology, particularly educational neuroscience. Educational neuroscience is an emerging field that incorporates methodologies both from cognitive and brain sciences and education [29]. The goal of educational neuroscience is not only to understand brain mechanisms for learning, cognition and affect, but also to study learning interventions [30]. Educational neuroscience targets developing interventions based on neurocognitive research, implements them in authentic contexts (e.g., schools) and draws conclusions about effective learning design practices. As part of this goal, educational neuroscience studies use behavioral and neuroimaging experiments, as well as design based research and implementation studies. Modeling is a useful tool in studying the long-term effects of interventions developed based on neurocognitive research. However, estimating long-term effectiveness of learning interventions and studying the sequencing and pacing of interventions using modeling are not well explored in education. As such, this study constitutes an example for how modeling approaches widely used in other fields (e.g., evolutionary biology, computer



science, social sciences) can be used to extrapolate results from shorter term interventions to build an understanding about how longer term interventions should be designed, and what outcomes can be expected.

Since educational neuroscience is a bourgeoning field the grand questions, methodologies and scientific communities for educational neuroscience have yet to consolidate. Having a foot both in cognitive and brain sciences, and education there are a wide range of methodologies, both quantitative and qualitative, available to use for educational neuroscientists. We group efforts in educational neuroscience into three categories: (1) Focus on mechanisms; these studies test mechanistic theories about cognitive and affective processes using experimental designs (i.e., behavioral and neuroimaging) in lab environments. This form of inquiry lacks ecological validity (lack of authentic contexts and tasks) due to constraints imposed by the experimental designs, however they provide generalizable results about cognitive and affective processes. (2) Intervention studies; these studies involve designing and studying the outcomes of an intervention. The design decisions for an intervention are informed by previous empirical and theoretical research, and the intervention can be thought as the operationalized form of the theory tested. Methodologies can be varied, including but not limited to experimental, quasi-experimental, and design-based studies. Interventions in educational research usually take place in authentic contexts and, therefore, complement the decontextualized lab studies. While lab studies inform mechanistic theories, intervention studies explicate how learning interventions justified by these theories hold their ground in authentic contexts. (3) Modeling and meta-analysis; Mathematical and computational modeling is a widely used method in cognitive science [31]. A model is essentially a formal and mathematical representation of a theory. In addition to being a formal representation for a theory, a computational model can also predict a



phenomenon (learning outcomes in this case). The comparison of this prediction with empirical results informs the validity of the model. Meta-analysis studies involve comparison of a number of studies that address a specific question to explore the wider or more generalizable insights gained. Both modeling and meta-analysis studies allow for extrapolation and further generalization of results gained from experimental and intervention studies.

*<Place Figure 6 Here>*

The three-fold model presented here does not provide a mutually exclusive way of categorizing different studies. Most studies involve aspects of these three categories. In addition, the ordering of the efforts in a research study can be varied. An intervention or a meta-analysis can trigger ideas about testing a particular mechanistic claim, or a lab study can implicate a specific intervention. Efforts under these three categories are cyclical, can follow any order, and are sometimes iterative (e.g., lab study yields to an intervention, intervention study unfolds further mechanistic questions).

This study responds to a previous call for the need to bridge affective neuroscience with educational design [32], and realizes the modeling aspect of a project that involved both behavioral and neuroimaging studies [23,33–37] on how different forms of moral exemplars modulated voluntary service behavior of a group of college students. In educational studies, due to practical time constraints, short-term implementation of an intervention usually does not produce the whole set of outcomes targeted. For example, a two-week implementation of an intervention at a school environment might show promising outcomes, yet not reach the ultimate expected threshold, but imply a trend towards it. In situations like this it is challenging to estimate the intervention durations and pacing that would lead to the desired outcomes. Modeling



and simulation approaches can provide the analytical tools to extrapolate results from the shorter-term interventions to inform long-term use of the interventions.

**Limitations**

However, there are several limitations in the present study that should be addressed by future studies. First, this ECM and Markov Chain-applied simulation model assumes homogeneity across individuals without considering intra- and inter-individual differences, such as differences in gender, SES and other demographical factors, and interactions between the whole system and individuals. Furthermore, the current simulation model assumes that the future trend will follow the current trend without considering the dynamic nature of human development [38–40]. Instead, it predicts future outcomes using causal matrices based on past states through iterations and extrapolations. To address these methodological shortcomings, additional advanced computation techniques, such as advanced machine learning algorithms [41], should be employed in future studies. Recently, it was shown that the knowledge of a student can be reliably tracked with a deep recurrent neural network [42]. Second, although we showed that our ECM model was able to predict long-term outcomes the results should be interpreted carefully because the prediction is easier to make when there are no interventions. Third, the current simulation model itself is only applicable to the presented case of moral exemplar interventions, and cannot be directly applied to other types of interventions in its current form. To address this issue, we plan to develop a GUI that will enable users to easily create their own ECM and Markov Chain by entering their intervention experiment data and to simulate longitudinal outcomes of their interventions with a customized model. Fourth, larger data should be used for the model building to enhance the generalizability of the simulation model. Because we only used a relatively small dataset ($n = 238$) collected from a single experiment, future studies should



employ larger datasets collected from other areas of psychology and educational studies to establish a more reliable and robust simulation methodology.

## Conclusions

The current simulation model will be able to particularly contribute to the following types of research in the field. First, the findings from the simulation can help intervention researchers who intend to conduct long-term, large-scale intervention experiments by being informed by short-term, small-scale pilot data in order to establish hypotheses; the simulation may be able to save time and resources. For instance, if researchers want to compare the influences of interventions between different frequencies, the findings from the current simulation model can inform them which frequencies they should test. On the one hand, in case of the moral exemplar intervention experiments, which were simulated in the present study, without such simulation results, researchers may have to test more than a couple of frequencies ranging from once per month to a couple of years. On the other hand, researchers who performed this simulation can have preliminary information regarding frequencies that were predicted to produce certain aimed effect sizes and may only have to conduct experiments testing those specific frequencies (e.g., once per 10.5 months [large effect] and 24 months [medium effect, statistical significance]). Second, this simulation model can provide guidelines for interventions to practitioners and policy makers. Because time and resources available in classroom settings are limited, newly invented educational programs should be applied in those settings with discretion [43]. For instance, if a principal intends to employ the moral exemplar intervention in her school, she may have to reduce time and resources already allocated to other subjects, such as mathematics and science, which are perhaps also fundamental for adolescents' academic achievement and career. If she has information regarding what an effective intervention type is (i.e., attainable and



relevant exemplar intervention) and the minimum required frequency of intervention producing a large effect (i.e., once per 10.5 months), which were presented by the current simulation model, she will be able to allocate class hours to this intervention program more effectively.

Consequently, despite the limitations that we have discussed, the simulation model described in the present study can contribute to the intervention methodologies in psychology and educational neuroscience, as well as providing applicable tools for educational programs focusing on adolescents.

## Acknowledgement

This research did not receive any specific grant from funding agencies in the public, commercial, or not-for-profit sectors. The authors thank Kelsie J. Dawson for her constructive comments.



Running head: PREDICTING EDUCATIONAL INTERVENTION OUTCOMES

Running head: PREDICTING EDUCATIONAL INTERVENTION OUTCOMES

**Tables**

|  |  | Conformers ($t$) | Non-conformers ($t$) |
|---|---|---|---|
| Cultural system 1 (Well organized, enough resources) | Conformers ($t+1$) | .80 (ECM [1,1,1]) | .60 (ECM [1,1,2]) |
|  | Non-conformers ($t+1$) | .20 (ECM [1,2,1]) | .40 (ECM [1,2,2]) |
| Cultural system 2 (Not organized, not enough resources) | Conformers ($t+1$) | .30 (ECM [2,1,1]) | .10 (ECM [2,1,2]) |
|  | Non-conformers ($t+1$) | .70 (ECM [2,2,1]) | .90 (ECM (2,2,2]) |

Table 1. Sample ECM for two different hypothetical cultural systems



| Study | Subjects | Participants | | Non-participants | | Total |
|---|---|---|---|---|---|---|
| Youniss et al. (2001) | High school juniors and seniors | 156 | (40.10%) | 223 | (57.33%) | 389 |
| Thoits et al. (2001) | Adults who were 25 years or older | 2,173 | (60.08%) | 1,444 | (39.92%) | 3,617 |
| | Adults who were 25 years or older | 1,642 | (57.27%) | 1,225 | (42.73%) | 2,867 |
| Hart et al. (2007) | 12th graders | 9,720 | (54.00%) | 8,280 | (46.00%) | 18,000 (approx.) |
| Total | | 13,691 | (55.07%) | 11,172 | (44.93%) | 24,863 |

Table 2. Number of service participants and non-participants extracted from large-scale survey data



| Data | Participants | | Non-participants | | Total |
|------|---------|---------|---------|---------|-------|
| Simulation | 120.86 | (50.78%) | 117.14 | (49.22%) | 238 |
| Real survey data | 13,691 | (55.07%) | 11,172 | (44.93%) | 24,863 |

Table 3. Comparison between the simulated result and real survey data



**Figures**

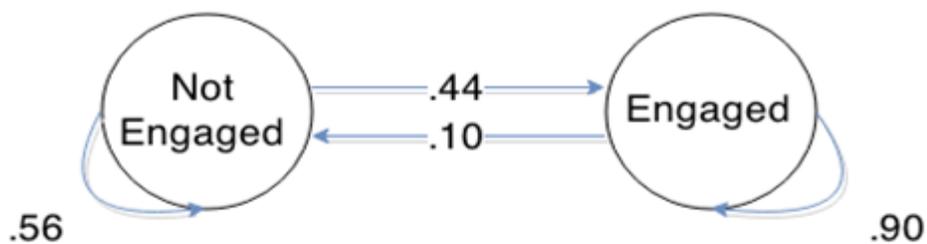

## (a) Attainable and relevant exemplar intervention

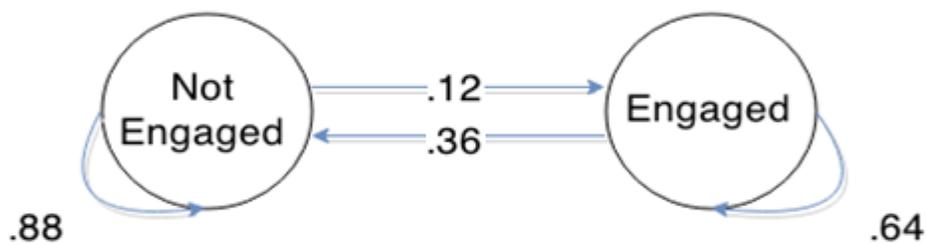

## (b) Extraordinary exemplar intervention

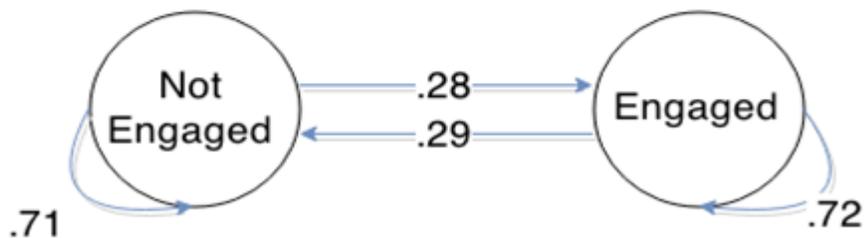

## (c) Without any intervention (control condition)

Figure 1. Created Markov Chains for three conditions



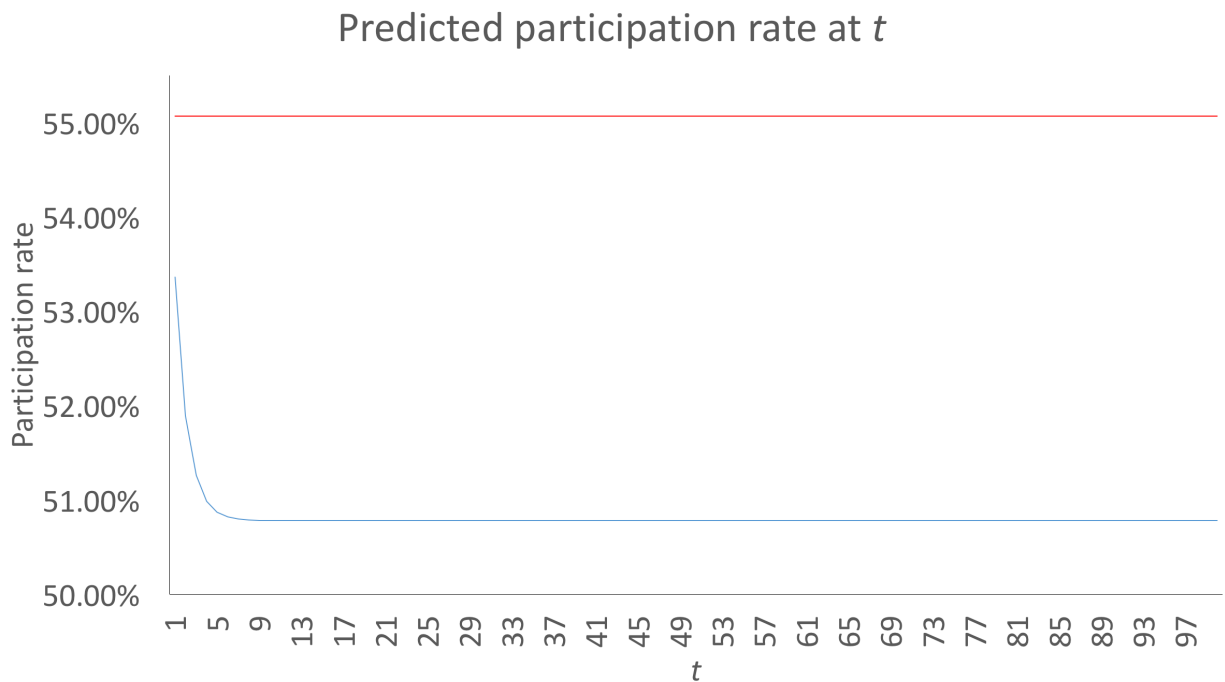

Figure 2. Predicted participation rate at *t* (blue line) and mean participation rate calculated from

the national survey results (red line)



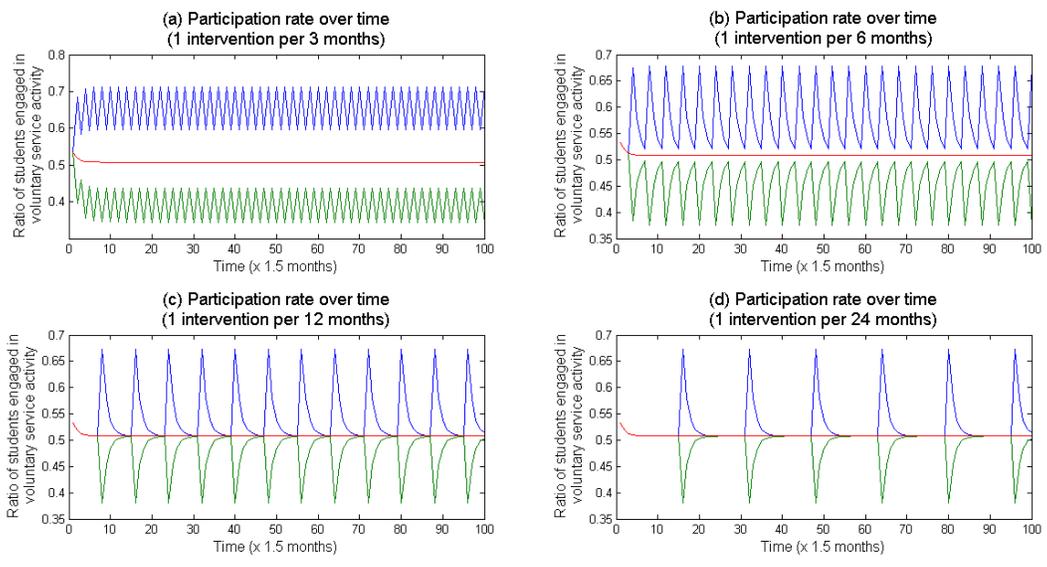

Figure 3. Different longitudinal trajectories of participation rate according to different intervention conditions and frequencies. Blue line: attainable and relevant exemplar intervention condition. Red line: control condition (baseline). Green line: extraordinary exemplar intervention condition.



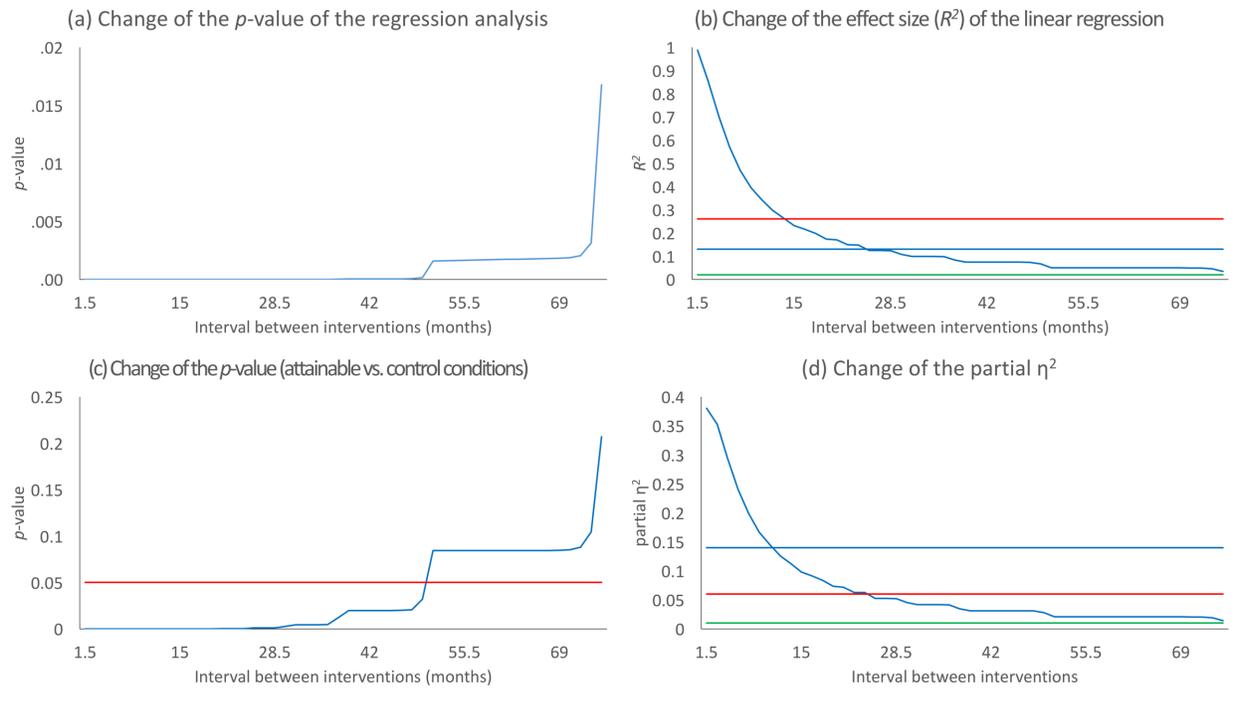

Figure 4. Results of statistical analyses. Red line: threshold for a large effect size ($R^2 = .26$ or partial $\eta^2 = .14$) or a significant difference ($p < .05$, Bonferroni's correction applied). Blue line: threshold for a medium effect size ($R^2 = .13$ or partial $\eta^2 = .06$). Greene line: threshold for a small effect size ($R^2 = .02$ or partial $\eta^2 = .01$).



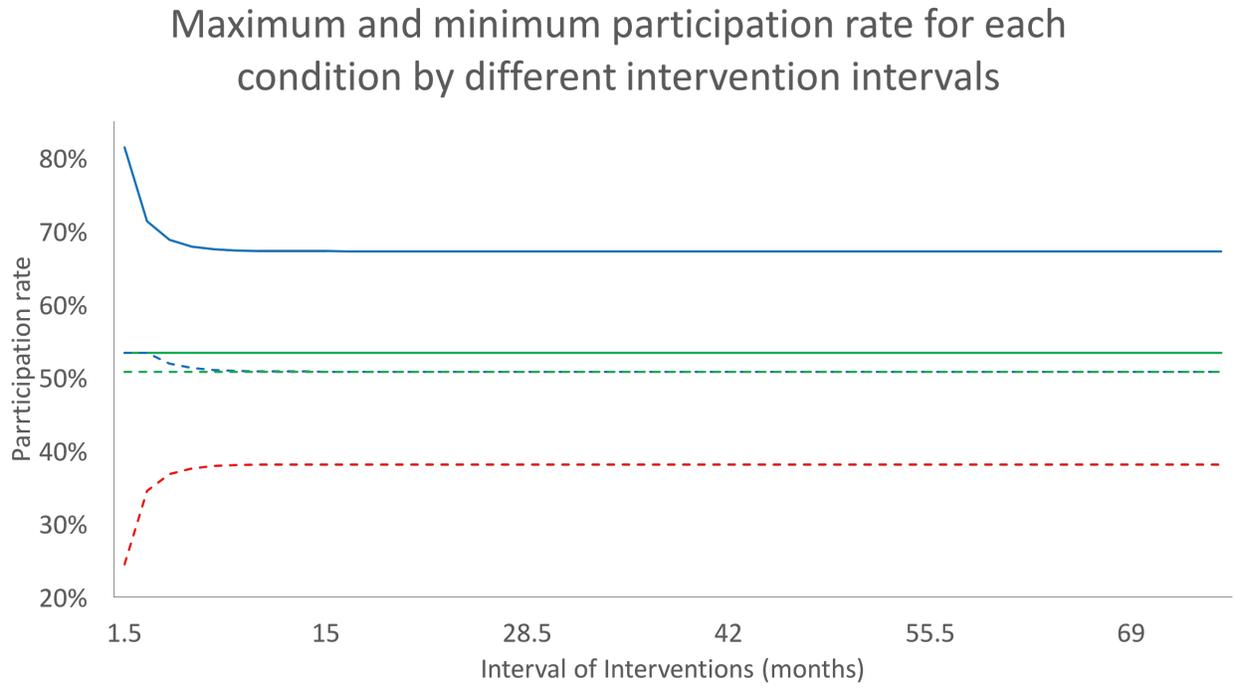

Figure 5. Predicted maximum and minimum participation rate for each condition and for each intervention frequency. Blue solid line: attainable and relevant exemplar condition maximum. Red solid line: unattainable and irrelevant exemplar condition maximum. Green solid line: control condition maximum. Blue dash line: attainable and relevant exemplar condition minimum. Red dash line: unattainable and irrelevant exemplar condition minimum. Green dash line: control condition minimum.



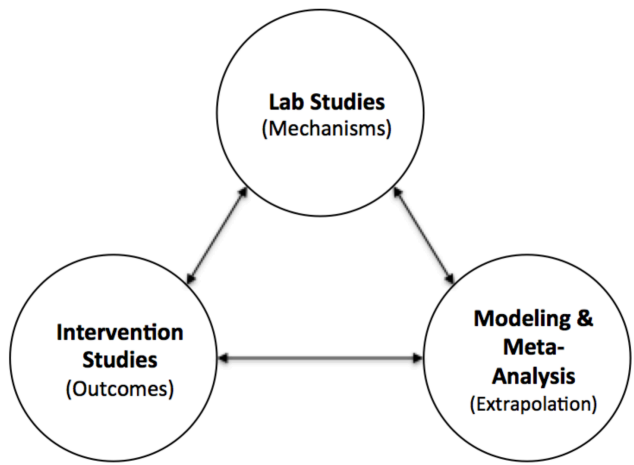

Figure 6. Three aspects of educational neuroscience research.



Running head: PREDICTING EDUCATIONAL INTERVENTION OUTCOMES

```
% by HH, KL and FS

% This MATLAB code simulates future outcomes of educational interventions

% as published in...

% Clear workspace and figure panels. Also clear variables.

clear;

clf;

figure(1);

record = 0;

% Set initial state

% ic = non participants

% it = participants

ic = 111;

it = 127;

% Initialize matrices
```



```
UC = zeros(50,100);

UT = zeros(50,100);

AC = zeros(50,100);

AT = zeros(50,100);

dataoutput = zeros (1,1);

% Define ECMs (Markov chain constants)

% for Attainable relevant exemplar intervention condition

P_AI = [18/32. 4/40; 14/32, 36/40];

% for Extraordinary exemplar intervention condition

P_UI = [30/34, 12/33; 4/34, 21/33];

% Without any intervetion

P_C = [32/45, 14/50; 13/45, 36/50];

% Extraordinary exemplar intervention condition.

% For different frequencies... 1/1.5months - 1/75months

for modu=1:50
```



```
% Set initial state again. For each frequency

UC(modu,1) = ic;

UT(modu,1) = it;

% From t = 1 to 100, iterative calculation.

for i=2:100

    % Whether it is "t" for intervention?

    if (mod(i,modu) == 0)

        % Yes.

        % Calculate states at t+1 based on t and the defined Markov

        % Chain in this condition.

        UC(modu,i) = UC(modu,i-1)*P_UI(1,1) +UT(modu,i-1)*P_UI(1,2);

        UT(modu,i) = UC(modu,i-1)*P_UI(2,1) +UT(modu,i-1)*P_UI(2,2);

    else

        % No intervention at this t
```



```
    % Calculate states at t+1 based on t and the defined Markov

    % Chain in the control condition.

    UC(modu,i) = UC(modu,i-1)* P_C(1,1) + UT(modu, i-1) * P_C(1,2);

    UT(modu,i) = UC(modu,i-1)* P_C(2,1) + UT(modu, i-1) * P_C(2,2);

  end

  % Store current states in memory

  record = record + 1;

  dataoutput(record,1) = 1;

  dataoutput(record,2) = modu;

  dataoutput(record,3) = i-1;

  dataoutput(record,4) = UT(modu,i)/(ic+it);

end

% From t = 1 to 100, calculate ratio of each state among the whole

% population

for i=1:100
```



```
        RC(i) = UC(modu,i) / (UC(modu,i)+UT(modu,i));

        RT(i) = UT(modu,i) / (UC(modu,i)+UT(modu,i));

    end

    % Calculate mean and std for all ts in this condition.

    UMT(modu) = mean(RT);

    UST(modu) = std(RT);

    UMC(modu) = mean(RC);

    USC(modu) = std(RC);

end

% Attainable-relevant exemplar intervention condition.

% For different frequencies... 1/1.5months - 1/75months

for modu=1:50

    % Set initial state again. For each frequency

    AC(modu,1) = ic;
```





```
AT(modu,1) = it;

% From t = 1 to 100, iterative calculation.

for i=2:100

    % Whether it is "t" for intervention?

    if (mod(i,modu) == 0)

        % Yes.

        % Calculate states at t+1 based on t and the defined Markov

        % Chain in this condition.

        AC(modu,i) = AC(modu,i-1)* P_AI(1,1) + AT(modu,i-1)* P_AI(1,2);

        AT(modu,i) = AC(modu,i-1)* P_AI(2,1) + AT(modu,i-1)* P_AI(2,2);

    else

        % No intervention at this t

        % Calculate states at t+1 based on t and the defined Markov

        % Chain in the control condition.

        AC(modu,i) = AC(modu,i-1)* P_C(1,1) + AT(modu, i-1) * P_C(1,2);
```



```
        AT(modu,i) = AC(modu,i-1)* P_C(2,1) + AT(modu, i-1) * P_C(2,2);

    end

    % Store current states in memory

    record = record + 1;

    dataoutput(record,1) = 2;

    dataoutput(record,2) = modu;

    dataoutput(record,3) = i-1;

    dataoutput(record,4) = AT(modu,i)/(ic+it);

end

% From t = 1 to 100, calculate ratio of each state among the whole

% population

for i=1:100

    RC(i) = AC(modu,i) / (AC(modu,i)+AT(modu,i));

    RT(i) = AT(modu,i) / (AC(modu,i)+AT(modu,i));

end
```



```
    % Calculate mean and std for all ts in this condition.

    AMT(modu) = mean(RT);

    AST(modu) = std(RT);

    AMC(modu) = mean(RC);

    ASC(modu) = std(RC);

end

% Set initial states for the control condition.

% One-dimensional array is sufficient since we do not have to consider

% different internvention frequencies in the control condition.

C(1) = ic;

TT(1) = it;

% From t = 1 to 100, iterative calculation.

for i=2:100

    % Calculate states at t+1 based on t and the defined Markov Chain in
```



Running head: PREDICTING EDUCATIONAL INTERVENTION OUTCOMES

```
    % the control condition.

    C(i) = C(i-1)* P_C(1,1) + TT( i-1) * P_C(1,2);

    TT(i) = C(i-1)* P_C(2,1) + TT( i-1) * P_C(2,2);

end

% From t = 1 to 100, calculate ratio of each state among the whole

% population

for i=1:100

    RC(i) = C(i) / (C(i)+TT(i));

    RT(i) = TT(i) / (C(i)+TT(i));

end

% Calculate mean and std for all ts in this condition.

tMT = mean(RT);

tST = std(RT);

tMC = mean(RC);

tSC = std(RC);
```





```
CMT(1:50) = tMT;

CST(1:50) = tST;

CMC(1:50) = tMC;

CSC(1:50) = tSC;

% Prepare for plots

x=2:50;

% First figure: Different mean participation ratio (throught t=0 to 100) by

% different conditions and by different frequencies

plot(x,AMT(2:50),x,UMT(2:50),x, CMT(2:50));

set(gca,'XTick',2:24:50) ;

set(gca,'XTickLabel',{'once per 1.5 months','per 37.5 months', 'per 75 months'},'FontSize',12) ;

xlabel('Frequency of interventions','FontSize',14);

ylabel('Mean ratio total engaged / total population','FontSize',14);

title('Mean ratio of students engaged in voluntary service activity','FontSize',14);

legend('Attainable exemplar intervention','Unattainable exemplar intervention','No intervention');
```



```
% Second fiture: Different Trajectories of participation ratio by different

% conditions and by different frequencies

figure(2);

x=1:100;

% Subplot 1: Trends when interventions were conducted 1 per 3 months

subplot(2,2,1);

plot(x,AT(2,:)/(ic+it),x,UT(2,:)/(ic+it),x,TT(:)/(ic+it));

xlabel('Time (x 1.5 months)','FontSize',12);

ylabel({['Ratio of students engaged in'], ['voluntary service activity']},'FontSize',12);

title({['(a) Participation rate over time'],['(1 intervention per 3 months)']},'FontSize',14);

% Subplot 2: Trends when interventions were conducted 1 per 6 months

subplot(2,2,2);

x=1:100;

plot(x,AT(4,:)/(ic+it),x,UT(4,:)/(ic+it),x,TT(:)/(ic+it));
```



```
xlabel('Time (x 1.5 months)','FontSize',12);

ylabel({['Ratio of students engaged in'], ['voluntary service activity']},'FontSize',12);

title({['(b) Participation rate over time'],['(1 intervention per 6 months)']},'FontSize',14);

% Subplot 3: Trends when interventions were conducted 1 per 12 months

subplot(2,2,3);

x=1:100;

plot(x,AT(8,:)/(ic+it),x,UT(8,:)/(ic+it),x,TT(:)/(ic+it));

xlabel('Time (x 1.5 months)','FontSize',12);

ylabel({['Ratio of students engaged in'], ['voluntary service activity']},'FontSize',12);

title({['(c) Participation rate over time'],['(1 intervention per 12 months)']},'FontSize',14);

% Subplot 4: Trends when interventions were conducted 1 per 24 months

subplot(2,2,4);

x=1:100;

plot(x,AT(16,:)/(ic+it),x,UT(16,:)/(ic+it),x,TT(:)/(ic+it));

xlabel('Time (x 1.5 months)','FontSize',12);
```



```
ylabel({['Ratio of students engaged in'], ['voluntary service activity']},'FontSize',12);

title({['(d) Participation rate over time'],['(1 intervention per 24 months)']},'FontSize',14);

% Now, perform ANOVA and series of comparisons.

% ANOVA: whether there was a significant effect of intervention type for

% each frequency?

% Comparisons (with Bonferroni's correction): whether attainable and

% relevant exemplar intervention significantly better promoted

% participation compared to the control condition with a certain frequency?

for i=2:50

    % make table

    table = [AT(i,:);TT;UT(i,:)];

    table = table.';

    %perform ANOVA

    [p,tbl,stats]=anova1(table,{'AT','C','UT'},'off');
```





```
%calculate eta2

eta2(i-1) = cell2mat(tbl(2,2))/cell2mat(tbl(4,2));

pvalue(i-1) = p;

% calculate Cohen's D between two conditions

ma = mean(AT(i,:));

sa = std(AT(i,:));

mc = mean(TT);

sc = std(TT);

compc =  sqrt((power(sa,2.0)*100 + power(sc,2.0)* 100) / 200);

Deff(i-1) = (ma-mc) / compc;

% Compare Effect of attainable and relevent exemplar intervention

[h, pt(i-1)] = ttest2(AT(i,:),TT(:));
```



```
% Bonferroni correction

pt(i-1) = 1- power(1-pt(i-1),3);

for j=1:100

    % write records for control condition

    record = record + 1;

    dataoutput(record,1) = 0;

    dataoutput(record,2) = i;

    dataoutput(record,3) = j-1;

    dataoutput(record,4) = TT(i)/(ic+it);

end

end

% plot statistical analysis results

% Subplot 1: ANOVA p-value for each frequency

figure(3);

subplot(2,2,1);
```





```
plot(2:50,pvalue);

xlabel('Frequency of interventions','FontSize',12);

ylabel('Significance (p-value)','FontSize',12);

title({['(a) Change of the p-value of the ANOVA result']},'FontSize',14);

% Subplot 2: ANOVA eta-squred value for each frequency. Red line: eta2 >

% .26 large effect size. Blue line: eta2 > .13 medium effect size. Greene

% line: eta2 > .02 small effect size

subplot(2,2,2);

plot(2:50,eta2);

xlabel('Frequency of interventions','FontSize',12);

ylabel('Effect size (eta2)','FontSize',12);

title({['(b) Change of the effect size (eta2) of the ANOVA result']},'FontSize',14);

hline=refline([0 .26]);

set(hline,'Color','r')

hline1=refline([0 .13]);

set(hline1,'Color','b')
```



```
hline2=refline([0 .02]);

set(hline2,'Color','g')

% Subplot 3: Comparisons between participation rate in the attainable and

% relevant exemplar intervention and control conditions. p-value after

% applying Bonferroni's correction. Red line: p < .05 threshold after

% applyig Bonferroni's correction.

subplot(2,2,3);

plot(2:50,pt);

xlabel('Frequency of interventions','FontSize',12);

ylabel('Significance (p-value)','FontSize',12);

title({['(c) Change of the p-value (Attainable vs. Control conditions)']},'FontSize',14);

hline=refline([0 .001]);

set(hline,'Color','r')

% Subplot 4: Same comparisons. Cohen's D effect size (attainable and

% relevant exemplar intervention condition - control condition). Red line:
```



```
% D > .8 large effect size. Blue line: D > .5 medium effect size. Green

% line: D > .2 small effect size.

subplot(2,2,4);

plot(2:50,Deff);

xlabel('Frequency of interventions','FontSize',12);

ylabel('Effect size (D)','FontSize',12);

title({['(d) Change of the effect size (D) of t-test']},'FontSize',14);

hline=refline([0 .8]);

set(hline,'Color','r')

hline1=refline([0 .5]);

set(hline1,'Color','b')

hline2=refline([0 .2]);

set(hline2,'Color','g')
```